\documentclass[prl,final,twocolumn,amsmath,amsfonts,amssymb,superscriptaddress]{revtex4}%
\usepackage{amssymb}
\usepackage{amsfonts}
\usepackage{amsmath}
\usepackage{subfigure}
\usepackage{graphicx}%
\providecommand{\U}[1]{\protect\rule{.1in}{.1in}}

\begin{document}
\preprint{HEP/123-qed}
\title[ ]{Quantum phase transition in many-flavor supersymmetric QED$_{3}$}
\author{Jorge G. Russo }
\affiliation{Instituci\'o Catalana de Recerca i Estudis Avan\c{c}ats (ICREA), Pg. Lluis
Compayns 23 - 08010 Barcelona, Spain,}
\affiliation{Departament de F\' \i sica Cu\' antica i Astrof\'\i sica and ICC, Universitat
de Barcelona - Mart\'\i\ Franqu\`es, 1, 08028 Barcelona, Spain.}
\author{Miguel Tierz }
\affiliation{Departamento de Matem\'{a}tica, Grupo de F\'{\i}sica Matem\'{a}tica, Faculdade
de Ci\^{e}ncias, Universidade de Lisboa, Campo Grande, Edif\'{\i}cio C6,
1749-016 Lisboa, Portugal.}
\pacs{PACS number}

\begin{abstract}
We study $\mathcal{N}=4$ supersymmetric QED in three dimensions, on a
three-sphere, with $2N$ massive hypermultiplets and
a Fayet-Iliopoulos parameter. We identify the exact
partition function of the theory with a conical (Mehler) function. This
implies a number of analytical formulas, including a recurrence relation and a
second-order differential equation, associated with an integrable system.
In the large $N$ limit,  the theory undergoes a second-order phase
transition on a critical line in the parameter space. We discuss the critical
behavior and compute the two-point correlation function of a gauge invariant
mass operator, which is shown to diverge as one approaches criticality from
the subcritical phase. Finally, we comment on the asymptotic $1/N$ expansion
and on mirror symmetry.

\end{abstract}
\volumeyear{year}
\volumenumber{number}
\issuenumber{number}
\eid{identifier}
\date[Date text]{date}
\received[Received text]{date}

\revised[Revised text]{date}

\accepted[Accepted text]{date}

\published[Published text]{date}

\startpage{101}
\endpage{102}
\maketitle


The study of 
 quantum electrodynamics in three dimensions (QED$_{3}$) has been a subject of interest since the early 1980s, due to its connection to finite-temperature QCD via dimensional reduction and the fact that, as QCD in four dimensions, QED$_{3}$ also exhibits spontaneous chiral symmetry
breaking and confinement \cite{Appelquist:1981vg}.

The theory has experienced a remarkably renewed interest in recent years due
to, in great part, the relevance of relativistic field theories of particles
moving in two dimensions in the description of the pseudogap phase of cuprates
\cite{Franz:2001zz}, the spin-liquid phase of quantum antiferromagnets \cite{Rantner:2002zz} and
the low-energy electronic excitations of graphene \cite{Kotov:2010yh}.
This reinvigorated relevance of the dynamics of a $U(1)$ gauge field and
$N_{f}$ fermions in three dimensions extends to the case with supersymmetry,
where the theory appears for example in descriptions of the physics of
half-filled Landau levels in terms of Dirac fermions \cite{Kachru:2015rma},
in  3d Bosonization \cite{Kachru:2016rui} and  nonperturbative descriptions of renormalization group flows \cite{Gukov:2016tnp}.
Supersymmetric $U(1)$ theories in three dimensions can be related to the study
of quantum phase transitions in quantum antiferromagnets and provide examples
of quantum phase transitions beyond the Landau-Ginzburg paradigm
\cite{Sachdev:2010zz}. In these discussions, the existence of mirror symmetry
in the supersymmetric gauge theory is important
\cite{Kachru:2016rui,Sachdev:2010zz} and it is precisely the recent, more
detailed, analysis of dualities that has bolstered great interest in QED$_{3}%
$. In particular for example, it has been recently shown that the fermionic
vortex of QED$_{3}$ is a free Dirac fermion
\cite{Metlitski:2015eka,Wang:2015qmt}. Around this result lurks a number of
connections between topological insulators, spin liquids and quantum Hall
physics, making QED$_{3}$ a subject of considerable physical interest.

At the same time, the development of tools for studying supersymmetric gauge
theories on curved manifolds, in particular localization
\cite{Pestun:2007rz,Pestun:2016zxk}, has increased the means at our disposal
to obtain exact analytical results. Examples
of works that use localization and the F-theorem in the study of QED$_{3}$
(with or without supersymmetry) includes Refs.
\cite{Klebanov:2011td,Giombi:2015haa,Dedushenko:2016jxl}. In this paper, we
shall study supersymmetric $U(1)$ gauge theory on $\mathbb{S}^3$, but, instead of massless matter as in Ref. \cite{Klebanov:2011td}, or
the cases studied in Refs. \cite{Gukov:2016tnp,Dedushenko:2016jxl}, we include  massive $\mathcal{N}$
$=4$ hypermultiplets and a Fayet-Iliopoulos (FI) term. As we
will see, this leads to a dramatic change in the dynamics of the theory. For
the sake of simplicity, we  consider the case of $N$ 
hypermultiplets with mass $m$ and $N$ 
hypermultiplets with
mass $-m$. More  general mass configurations will be discussed at the end. The total number of flavors $N_{f}$ is
therefore $2N$.
In addition, as mentioned, there is a FI term and it is actually the interplay between the difference of masses and the FI
parameter $\eta$ the one which is responsible for producing novel behavior. In
particular, it is responsible for the emergence of a second-order quantum
phase transition in the large $N$ limit.

We thus consider an $\mathcal{N}$ $=4$ supersymmetric $U(1)$
theory consisting of $2N$ massive $\mathcal{N}$ $=4$ (flavor) hypermultiplets
($N$ of mass $m$ and $N$ of mass $-m$), coupled to an $\mathcal{N}$ $=4$
vector multiplet. Localization readily leads to an integral representation for
the partition function \cite{Kapustin:2009kz}%
\begin{align}
Z_{\mathrm{QED}_{3}}  &  =\int_{-\infty}^{\infty}dx\ \frac{e^{i\eta x}%
}{\left[  2\cosh(\frac{x+m}{2})2\cosh(\frac{x-m}{2})\right]  ^{N}}\nonumber\\
&  =2^{-N}\int_{-\infty}^{\infty}dx\ \frac{e^{i\eta x}}{\left[  \cosh x+\cosh
m\right]  ^{N}}. \label{ZQED}%
\end{align}
In what follows, we drop the $2^{N}$ factor, which is inessential to our
discussion and we have  set the radius of $\mathbb{S}^3$ to $r=1/\left(
2\pi\right)  $. In the case when the hypermultiplets have masses $m_{1}$ and
$m_{2}$, the parameter $m$ is 
$m=\big(  m_{1}-m_{2}\big)  /2$. Thus, in
the discussion below, increasing $m$ corresponds to separating the two mass scales.
It is important to note that the parameters $m$, while they correspond to mass deformations of the Lagrangian, represent curved-space analogs
of the more familiar flat space mass parameters, as here they 
 are  measured in units of the radius of the $\mathbb{S}^3$ (the partition function
 only depends on the combination $mr$).

Notice that, by writing the integral representation  (\ref{ZQED}) in the
latter form, one can immediately identify it as the integral representation of
a conical (Mehler) function \cite{conicalref}, which is an associated Legendre
function with a complex index. We find
\begin{equation}
Z_{\mathrm{QED}_{3}} =\sqrt{2\pi}\ \frac{ \Gamma\left(  N+i\eta\right)
\Gamma\left(  N-i\eta\right)  }{\Gamma\left(  N\right)  (\sinh(m))^{N-\frac
{1}{2}} } \ P^{\frac{1}{2}-N}_{-\frac{1}{2}+i\eta} (\cosh(m))\ .
\end{equation}
It can also be conveniently represented in terms of a hypergeometric
function,
\begin{align}
Z_{\mathrm{QED}_{3}}  &  =\frac{\sqrt{2\pi}\ \Gamma\left(  N+i\eta\right)
\Gamma\left(  N-i\eta\right)  }{\Gamma\left(  N\right)  \Gamma\left( N+
\frac{1}{2}\right)  (1+z)^{N-\frac{1}{2}} }\label{dos}\\
\times &  \ {}_{2}F_{1}\Big(  \frac{1}{2}-i\eta,\frac{1}{2}+i\eta,N+\frac
{1}{2};\frac{1}{2} (1-z)\Big)  .\nonumber
\end{align}
with $z\equiv\cosh(m)$.
In specific cases, the expression simplifies. In particular, for two and four
flavors, we find
\begin{align}
&  Z_{\mathrm{QED}_{3}}^{N=1}=\frac{2\pi\sin\left(  m\eta\right)  }%
{\sinh\left(  m\right)  \sinh\left(  \pi\eta\right)  }, \label{n1}\\
&  Z_{\mathrm{QED}_{3}}^{N=2}=\frac{2\pi\left(  \cosh m\sin\left(
m\eta\right)  -\eta\sinh m\cos\left(  m\eta\right)  \right)  }{\sinh
^{3}\left(  m\right)  \sinh\left(  \pi\eta\right)  }.\label{n2}%
\end{align}
These can also be obtained from residue integration \cite{Benvenuti:2011ga}.
The expression (\ref{n1}) already exhibits some of the general properties
of the conical function and, hence, of the partition function, such as the
oscillatory behavior, which depends on both $m$ and $\eta$. It is well known
that supersymmetric QED$_{3}$ with two flavors is self-dual
\cite{Intriligator:1996ex}. Notice that indeed we find that (\ref{n1}) is
invariant under the exchange $m\leftrightarrow\pi\eta$ implied by the duality
transformation.

The oscillatory behavior is related to the fact that the function has an
infinite number of zeros, all of them real, precisely in the physical region
$m\geq0$ \cite{conicalref} and the function is monotonic
 until the appearance of the first zero of the function, after which
the behavior is oscillatory. This transition between a monotonic and an
oscillatory region when the first zero appears, in the large $N$ limit,
becomes a phase transition, which we characterize below by computing the
saddle points of (\ref{ZQED}).

 The identification with a conical function
and the ensuing hypergeometric representation has interesting consequences.
To begin with, from a standard
recurrence relation for the Legendre functions, we obtain that the partition
function also satisfies a recurrence relation:
\begin{align}
& (2N-1)\cosh(m)Z_{N}\\
& =\frac{(N-1)^{2}+\eta^{2}}{N-1}\ Z_{N-1}+N\sinh^{2}(m)\ Z_{N+1}\nonumber
\end{align}
For short, here we defined $Z_{N}\equiv Z_{\mathrm{QED}_{3}}(m,\eta,N)$. By
this formula, we can easily generate any $Z_{N}$ from the above expressions
(\ref{n1}) and (\ref{n2}) for $N=1$ and $N=2$. In addition, the representation (\ref{dos}) can be used to study a small mass expansion, since
the radius of convergence of a Gauss hypergeometric function is $\left\vert
x\right\vert <1$ in the variable. In the massless limit, the hypergeometric
becomes 1 and $Z_{\mathrm{QED}_{3}}$ is given by the first line in
(\ref{dos}). In order to find the large mass behavior, we use an Euler
hypergeometric transformation and write the partition in the form%
\begin{align}
Z_{\mathrm{QED}_{3}}  & =\frac{\Gamma(-i\eta)\Gamma(N+i\eta)\,}{2^{i\eta
}\Gamma(N)(\cosh(m)+1)^{N+i\eta}}\label{Zaltern}\\
& \times\text{ }_{2}F_{1}\Big(  \frac{1}{2}+i\eta,N+i\eta,1+2i\eta
;\mathrm{sech}^{2} \frac{m}{2}  \Big)  +{c.c.}\ \nonumber
\end{align}
Using (\ref{Zaltern}) we then obtain
\begin{equation}
Z_{\mathrm{QED}_{3}}^{mr\gg1}=\frac{2^{N}\Gamma(-i\eta)\Gamma(N+i\eta)}%
{\Gamma(N)}e^{-mr(N+i\eta)}+{c.c.}.\label{masaa}%
\end{equation}
Here, we have  restored the $\mathbb{S}^3$ radius dependence
to exhibit the fact that
this regime can also be interpreted as a decompactification limit.
Another nontrivial consequence of the relation of the partition function to a
hypergeometric function is the fact that then $Z_{N}$ satisfies a second-order
differential equation:
\begin{equation}
\frac{d^{2}Z_{N}}{dm^{2}}+2N\coth(m)\frac{dZ_{N}}{dm}\newline+\left(  \eta
^{2}+N^{2}\right)  Z_{N}=0\label{diffeq} .
\end{equation}
\ By defining $\widetilde{Z}_{\mathrm{QED}_{3}}=\left(  \sinh\left(  m\right)
\right)  ^{N}Z_{\mathrm{QED}_{3}}$, this equation can be written as a
Schr\"{o}dinger equation with a hyperbolic P\"{o}schl-Teller potential
\cite{Schindler}, which is a well-known solvable one-dimensional quantum mechanical problem,
\[
\frac{d^{2}\widetilde{Z}_{\mathrm{QED}_{3}}}{dm^{2}}+\left(  \eta^{2}%
+\frac{N(1-N)}{\sinh^{2}m}\right)  \widetilde{Z}_{\mathrm{QED}_{3}}=0.
\]


The present theory has a large $N$ limit, with fixed $\lambda\equiv\eta/N $.
In this limit, the partition function (\ref{ZQED}) can be computed by the
saddle-point method. The integrand in (\ref{ZQED}) can be written as
$e^{-NS(\lambda)}$ where the \textit{action }$S$ is
\[
S(\lambda,x,z)=-i\lambda x+\log(\cosh x+\cosh m)\ .
\]
The saddle-point equation is then
\[
-i\lambda+\frac{\sinh x}{\cosh x+\cosh m}=0,
\]
which has as solutions
\begin{equation}
x_{1,2} =\log\left(  \frac{-\lambda \cosh m\pm i\Delta}{i+\lambda}\right)  +2\pi in,
\end{equation}
where $n\in%
\mathbb{Z}
$ and  $\Delta\equiv\sqrt{1-\lambda^{2}\sinh^2m}$. In what follows we show that the theory undergoes a large $N$ phase
transition at $
\lambda_{c}\equiv 1/\sinh m,$
or, more generally, at the critical line $\lambda\sinh(m)=1$ in the $(\lambda,
m)$ space, where $\Delta=0$.

\noindent \emph{Subcritical phase} ($\lambda\sinh(m)<1$). In this case all saddle points
lie on the imaginary axis. We find that the saddle point $x_{1}$ with $n=0$ is
the relevant one, and, to leading order for large $N$, the partition function
becomes
\begin{equation}
Z_{\mathrm{QED}_{3}}\approx\frac{\sqrt{2\pi}}{\sqrt{NS^{\prime\prime}\left(
x_{1}\right)  }}\exp\left(  -NS\left(  x_{1}\right)  \right)  . \label{Zsub}%
\end{equation}
where
$S^{\prime\prime}(z,x)=\left(  z\cosh\left(  x\right)  + 1 \right)  /\left(
\cosh\left(  x\right)  +z \right)  ^{2}.$ We numerically checked that, for large $N$, this formula reproduces the analytic expression (\ref{dos}) with great accuracy. In the subcritical phase,
the large $N$ free energy $F=-\log Z_{\mathrm{QED}_{3}}$ is given by $N
S\left(  x_{1}\right)  $. We obtain
\[
F_{\mathrm{sub}}=-\frac{i\lambda N}{2}\log\! \left(  \frac{\left(  -z\lambda
+i\Delta\right)  \left(  i-\lambda\right)  }{\left(  z\lambda+i\Delta\right)
\left(  i+\lambda\right)  }\right)  +N\log\! \left(  \frac{z+\Delta}%
{1+\lambda^{2}}\right)  
\]

\noindent \emph{Supercritical phase} ($\lambda\sinh(m)>1$). 
The two saddle points move to the complex plane, with $x_2=-x_1^*$.
The action is complex, with $\operatorname{Re}\left(  S\left(  x_{1}\right)  \right)
=\operatorname{Re}\left(  S\left(  x_{2}\right)  \right)  ,$  $\operatorname{Im}\left(  S\left(
x_{1}\right)  \right)  =-\operatorname{Im}\left(  S\left(  x_{2}\right)
\right)  .$ Therefore, both saddle points contribute with equal weights and need to be taken into account.
The partition function $Z=Z_{\mathrm{QED}_{3}}$ is now%
\[
Z\approx\sqrt{\frac{2\pi}{N}}e^{-N\operatorname{Re}\left(  S\left(
x_{1}\right)  \right)  }\left(  \frac{e^{-iN\operatorname{Im}\left(  S\left(
x_{1}\right)  \right)  }}{\sqrt{S^{\prime\prime}\left(  x_{1}\right)  }}%
+c.c.
\right)  .
\]
For $N\gg1$, this expression  agrees with the exact
analytic expression (\ref{dos}). For large mass, it is also in precise
agreement with the large mass formula (\ref{masaa}). Taking the log to get the
free energy, we see that at large $N$, the leading contribution proportional
to $N$ is given by the real part of the action $F\approx N\operatorname{Re}%
\left(  S\left(  x_{1}\right)  \right)  $, giving%
\[
F_{\mathrm{super}}=\frac{N}{4}\left(  2\log\frac{z^{2}-1}{1+\lambda^{2}%
}-2i\lambda\log\frac{\lambda-i}{\lambda +i}\right)  .
\]
The free energy and its first derivative are continuous at the critical
point, while the second derivative gives
\begin{align*}
\frac{d^{2}F}{d\lambda^{2}} &  =\frac{N}{1+\lambda^{2}}\left(  1+\frac
{\cosh(m)}{\sqrt{1-\lambda^{2}\sinh^{2}(m)}}\right)  \ ,\quad\lambda
<\lambda_{c},\\
\frac{d^{2}F}{d\lambda^{2}} &  =\frac{N}{1+\lambda^{2}}\ ,\qquad\qquad
\qquad\qquad\qquad\qquad\quad\lambda\geq\lambda_{c}.
\end{align*}

\noindent Thus, $d^{2}F/d\lambda^2$ is discontinuous, implying a
second-order phase transition. In addition, in the subcritical regime the
susceptibility $\chi=-\frac{d^{2}F}{d\lambda^{2}}$ diverges as the critical
line is approached,
$\chi\sim(\lambda_{c}-\lambda)^{-\gamma}\ ,\ \gamma=1/2$,
which is a recurrent behavior in second-order phase transitions.  The critical
behavior is shown in Figs. 1(a) and 1(b) for fixed
$m=1$ and in Fig. 2 for the whole phase diagram.

\begin{figure}
\centering
\subfigure{\includegraphics[width=7.cm]{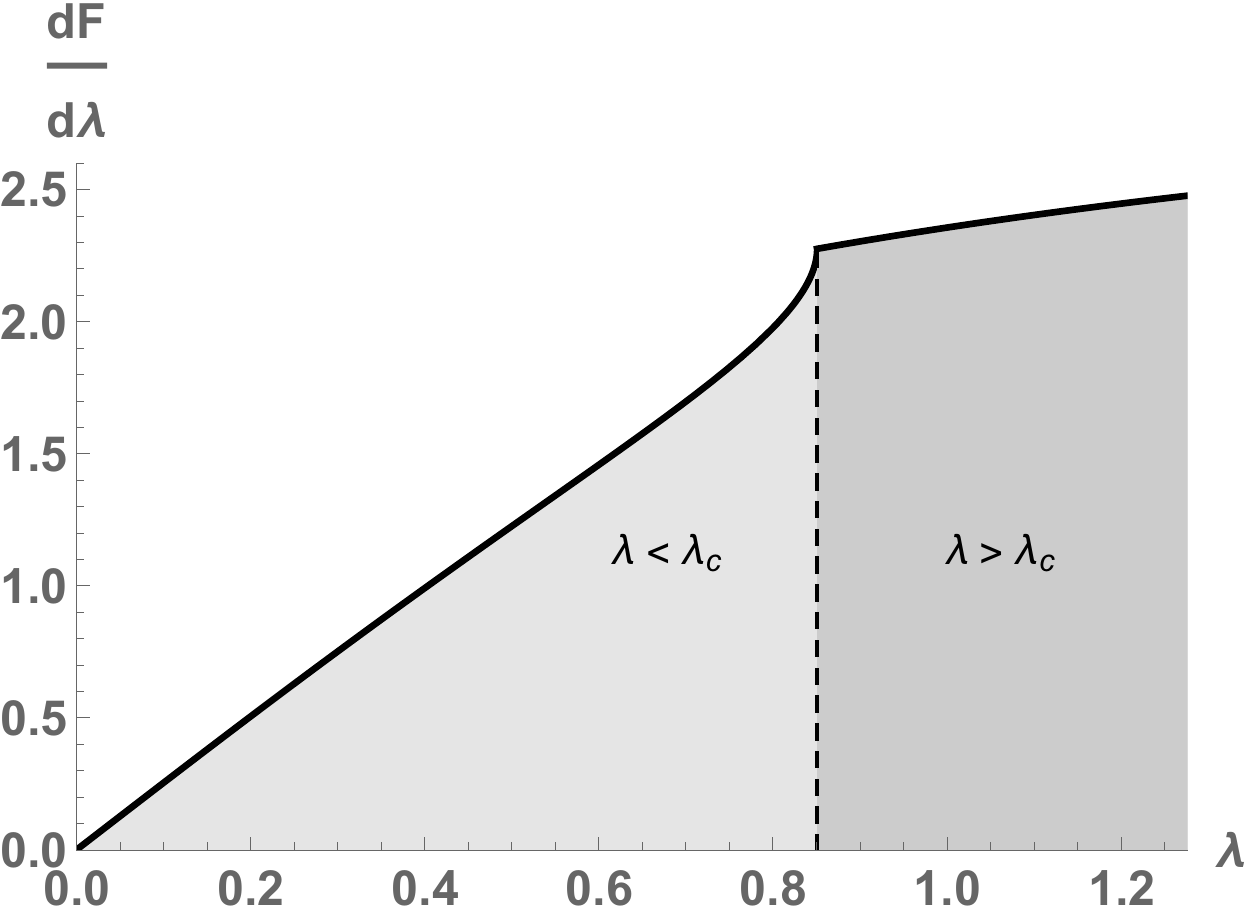}}
\ \ \ \ \subfigure{\includegraphics[width=7.cm]{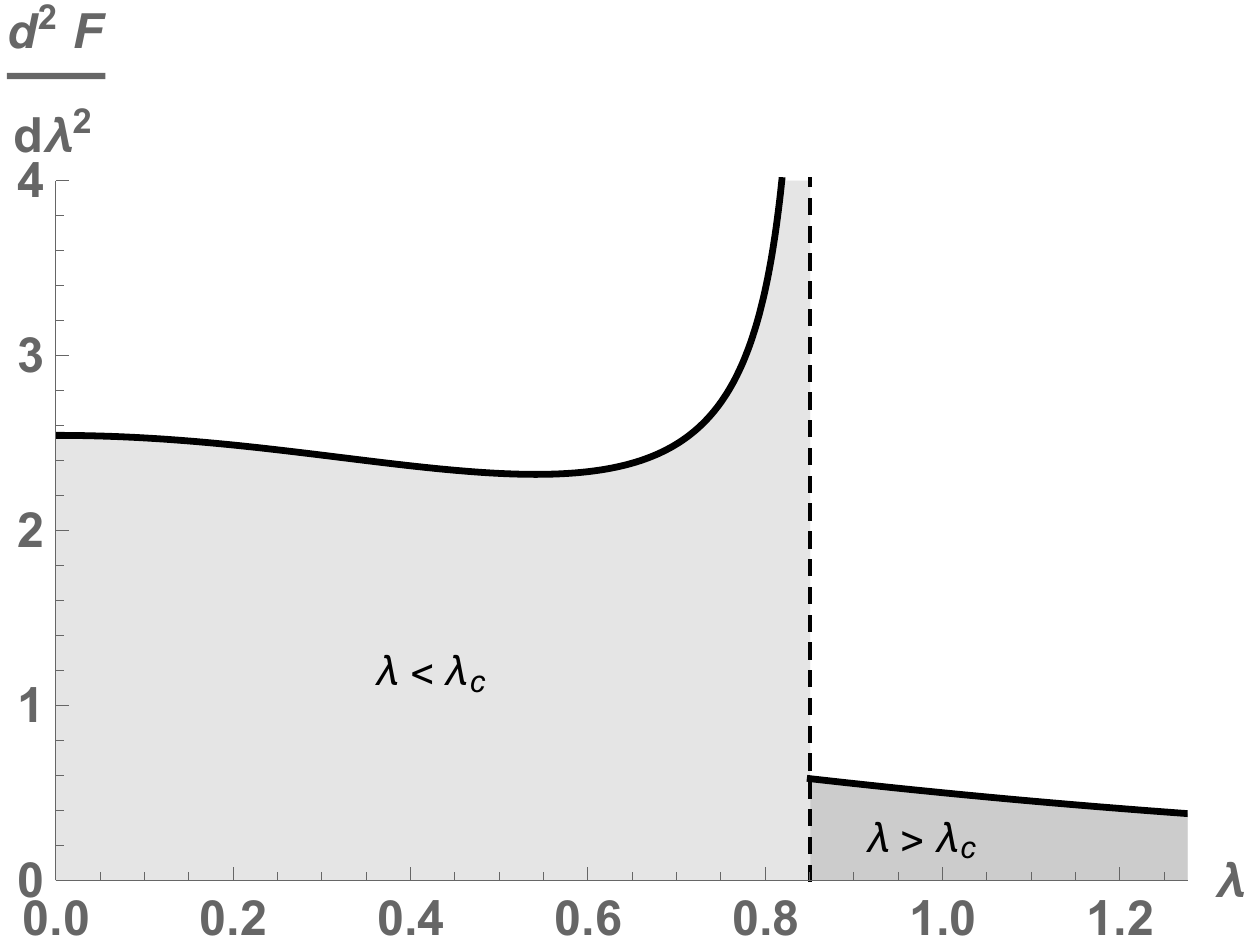}}\caption{(a) Behavior
of $dF/d\lambda $. (b) Discontinuity of $d^{2}F/d\lambda^{2}$ at the transition point ($m=1$). }%
\label{fig1charge}%
\end{figure}

\begin{figure}
\includegraphics[width=6.cm]{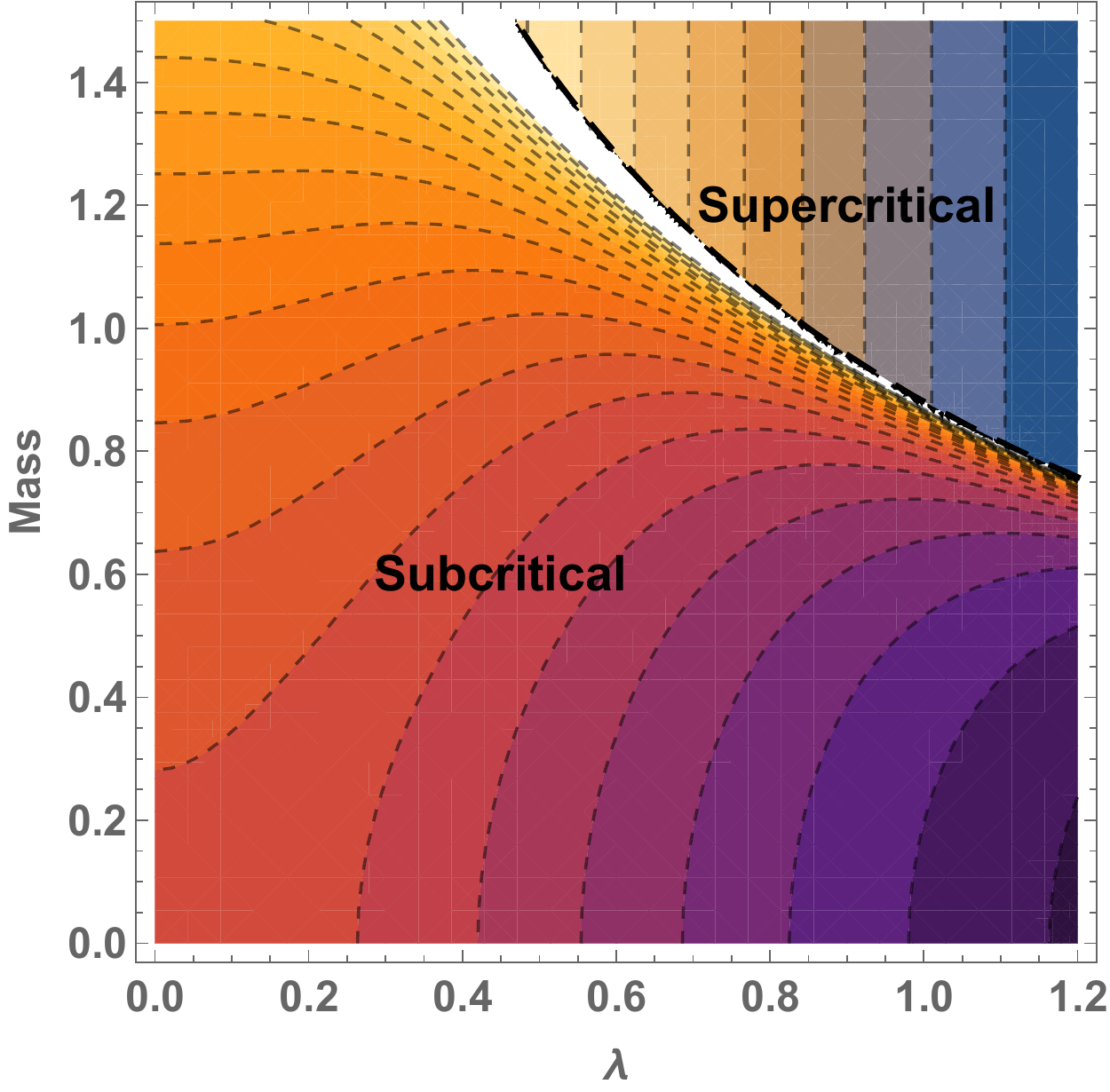} \caption{Phase diagram.
The critical line (dashed) $\lambda \sinh (m)=1$ separates the two  phases. The plot also shows the contour lines of
 $d^{2}F/d\lambda^{2}$ (increasing from dark to light).}
\end{figure}

Next, consider the analytic properties of the free energy in crossing the
critical line by varying the mass parameter at fixed coupling $\lambda$. By
differentiating the free energy with respect to the mass $m$, one generates
correlators of the gauge invariant mass operator \cite{Dedushenko:2016jxl}%
\[
J_{3}=\frac{1}{2N}\left(  \widetilde{Q}_{1,i}Q_{1}^{i}-\widetilde{Q}_{2,i}%
Q_{2}^{i}\right)  ,
\]
where $Q_{1}^i,\widetilde{Q}_{1i}$ are the  hypermultiplets of mass $m$ and $Q_{2}^i,\widetilde{Q}_{2i}$ are the  hypermultiplets of mass $-m$. Because of supersymmetry, these correlators are
independent of the position \cite{Dedushenko:2016jxl}. For example, for the
simple $N=1$ case, we have that%
\begin{align*}
\left\langle J_{3}\right\rangle  &  \varpropto\frac{d F}{d m}=\eta\cot\left(
m\eta\right)  -\coth\left(  m\right)  ,\\
\left\langle J_{3}J_{3}\right\rangle -\left\langle J_{3}\right\rangle
\left\langle J_{3}\right\rangle  &  \varpropto\frac{d^{2}F}{d m^{2}}%
=-\frac{\eta^{2}}{\sin^{2}\left(  m\eta\right)  }+\frac{1}{\sinh^2\left(
m\right)  }.
\end{align*}
This extends a result of Ref. \cite{Dedushenko:2016jxl} to the
case $\eta\neq0.$ Notice also that the differential equation (\ref{diffeq}) immediately establishes an exact relationship between $\left\langle J_{3}J_{3}\right\rangle$ and $\left\langle J_{3}\right\rangle$ for any finite $N$. 
Returning to the large $N$ free energy, we find that $\langle J_{3}\rangle $ is continuous, whereas 
\begin{align*}
\left(  \frac{d^{2}F}{dm^{2}}\right)  _{\lambda<\lambda_{c}}  &  =\frac
{1}{N\sinh^{2} m}\left( 1 -\frac{\cosh m}{\sqrt{1-\lambda^{2}\sinh^{2} m}%
}\right)  \\ 
\left(  \frac{d^{2}F}{dm^{2}}\right)  _{\lambda>\lambda_{c}}  &  =\frac
{1}{N\sinh^{2}m} \ .
\end{align*}
Thus, $d^{2}F/dm^{2}$ is discontinuous, implying a discontinuity in the
two-point function of the operator $J_{3}$. Moreover, the two-point
correlation function diverges as the critical line is approached from the
subcritical phase.

The theory has an asymptotic $1/N$ expansion, which we now briefly outline.
For concreteness, we consider the subcritical phase. 
The first $1/N$ correction arises from the term $(x-x_{1})^{4}$
in the expansion of  the integrand of (\ref{ZQED}) around $x_1$.
%
A
closely related expansion of the conical functions in inverse powers of
$(N-1/2)$ was discussed in Refs. \cite{Dunster, conicalref}.
An interesting approach is described in Ref. \cite{Gukov:2016tnp}.

An elegant treatment, which exhibits the asymptotic character of the $1/N$
series, is as follows. We introduce a new integration variable by the
transformation
\begin{equation}
S(x)-S_{0}=\beta t\ . \label{transf}%
\end{equation}
This leads to
\begin{equation}
Z_{\mathrm{QED}_{3}}=\beta e^{-NS_{0}}\int_{\mathcal{C}}dt\text{ }e^{-N\beta
t}\mathcal{B}\left(  t\right)   ,\text{  \ \ }\mathcal{B}\left(
t\right)  \equiv\frac{1}{S^{\prime}(x(t))} , \label{Laplace}%
\end{equation}
with $S_{0}\equiv S(x_{1}) ,\ \beta\equiv S^{\prime\prime}
(x_{1})/2$.
The contour $\mathcal{C}$ in the complex $t$-plane is determined by the
transformation (\ref{transf}) in varying $x$ from $-\infty$ to $\infty$
(see Fig. 3). The contour surrounds singularities at $t_{1}^{(n)}$ and
$t_{2}^{(n)}$ lying on the positive real axis, which are associated with the
saddles at $x_{1}$, with $n=0,1,2,....$, $x_{2}$, with $n=1,2,...$. All
singularities in $\mathcal{B}\left(  t\right)  $ are branch points of the form
$(t-t_{1,2}^{(n)})^{-1/2}$.
The $1/N$ expansion is generated upon Taylor expanding $\mathcal{B}\left(
t\right)  $ in powers of $t$,%
\[
\mathcal{B}\left(  t\right)  =\frac{1}{2\beta\sqrt{t}}\sum_{k=0}^{\infty}b_{k}
\ t^{k}\text{ }\ ,\qquad b_{0} =1\ .
\]
\begin{figure}[h]
\centering
\includegraphics[scale=.5]{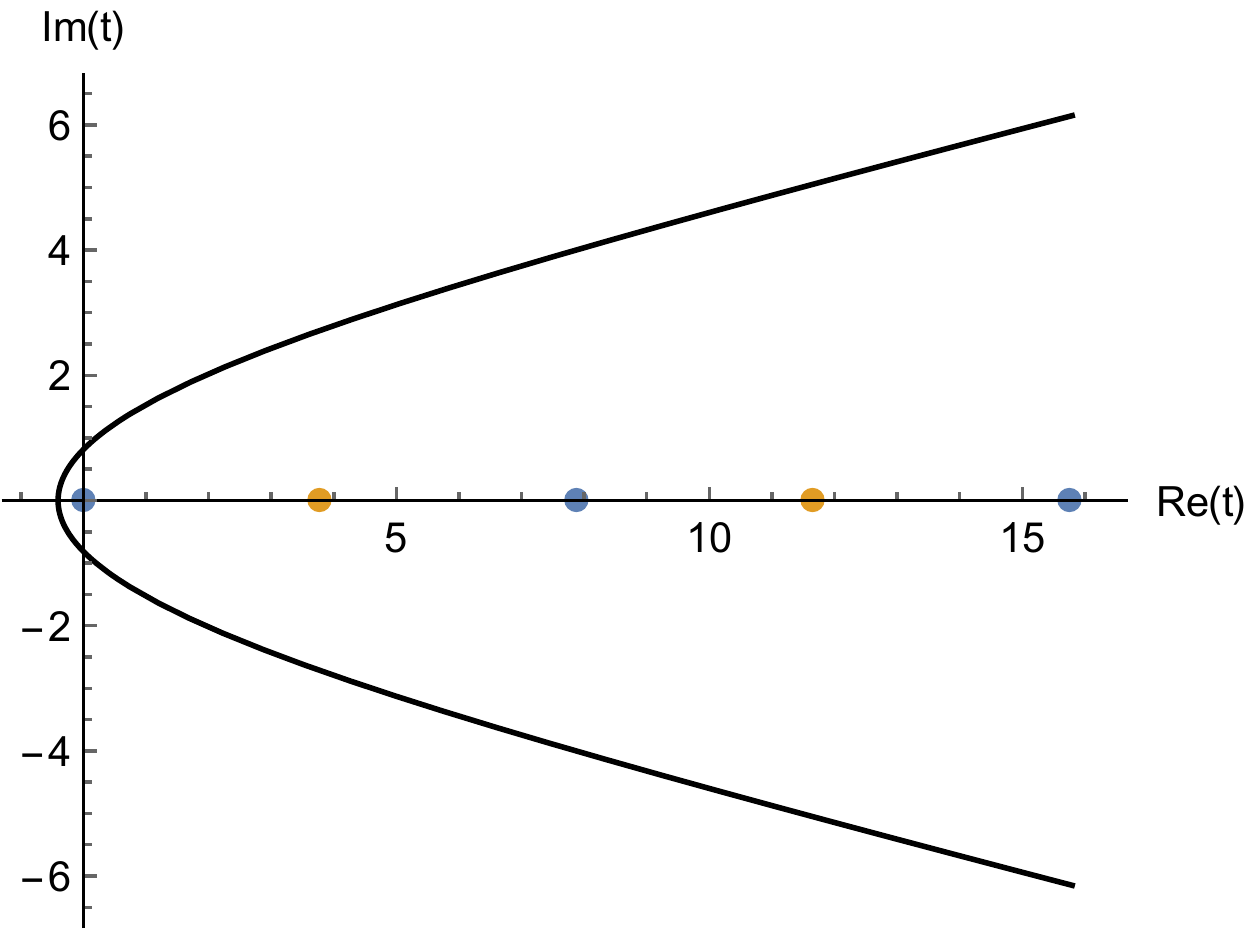}\caption{Integration contour
$\mathcal{C}$ in (\ref{Laplace}) and location of singularities of the
integrand ($m=1,\ \lambda=0.3 \lambda_{c}$). }%
\label{figu3}%
\end{figure}
This expansion has a finite radius of convergence, determined by the location
of the singularity that is closest to the origin. We are left with the
integral
\[
\int_{\mathcal{C}}dt\text{ }e^{-N\beta t}\ t^{k-\frac{1}{2}}=2\!\int_{0}%
^{\infty}dt\text{ }e^{-N\beta t}\ t^{k-\frac{1}{2}}=\frac{2\Gamma
(k+1/2)}{(\beta N)^{k+\frac{1}{2}}},%
\]
where we have deformed the contour to the positive real axis (note that in
this integral there is no singularity on the positive real axis). Thus we get
the asymptotic series%
\begin{equation}
Z_{\mathrm{QED}_{3}}=\frac{e^{-NS_{0}}}{(\beta N)^{\frac{1}{2}}}\sum
_{k=0}^{\infty}b_{k}\frac{\Gamma\left(  k+1/2\right)  }{\beta^{k} N^{k}}\ .
\label{series}%
\end{equation}
The term $k=0$ just  reproduces the earlier formula (\ref{Zsub}).

Here, we have expanded $\mathcal{B}\left(  t\right)  $ around $t=0$. By
expanding $\mathcal{B}\left(  t\right)  $ around some $t_{1,2}^{(n)}$,
$n=1,2,...$ one finds an extra factor $e^{-2\pi n \lambda}$ coming from
$e^{-NS(t)}$ . The presence of an infinite number of saddle points suggests
that the $1/N$ expansion can be more conveniently treated in terms of
resurgent trans-series. In deforming the contour, one crosses Stokes
discontinuities which may imply resurgent relations in the different
trans-series coefficients (see Refs.
\cite{Pasquetti:2009jg,Marino,Dunne:2012ae,Aniceto:2014hoa,Couso-Santamaria:2015wga} for
examples). It would be extremely interesting to understand the origin of
nonperturbative effects that render the large $N$ expansion asymptotic, as
well as the resurgent properties of the series and how the existence of a
phase transition is encoded in the $1/N$ expansions below and above the phase transition.

Now consider more general masses.  The theory with $N$ hypermultiplets of mass $m_{1}$ and 
$N$ hypermultiplets of mass $m_{2}$ is  equivalent to
the one we discussed.
By a shift in the integration variable, one gets
the same partition function (\ref{ZQED}) with an extra phase
$e^{-i\eta{m_{+}}}$ and $m$ replaced by $m_-$, with
$m_{\pm}\! =\! \left(  m_{1}\! \pm \! m_{2}\right)  /2$.
In general, for $N_f$ flavors, there are $N_f-1$ mass parameters associated with the Cartan generators of  $SU(N_f)$ flavor symmetry, satisfying 
$\sum_i m_i =0$. One can have
 independent parameters $m_1$ and $m_2$  by adding an extra hypermultiplet of mass $m_3=-N(m_1+m_2)$.
At large $N$, this decouples (its one-loop partition function
becomes a constant, $1/\cosh m_3$), and the 
large $N$ physics is then the same as in (\ref{ZQED}).
 More general mass assignments with  similar phase transition are  possible. The reason is that the mechanism that triggers the phase transition is also at work in more general cases: on the imaginary axis, the one-loop partition function provides a periodic potential with infinite number of vacua; as the
 constant force, represented by the FI parameter, is increased, there is  a critical point where this  overcomes the maximum force from the periodic potential. Beyond this point, equilibrium is not possible and the saddle points move to the complex plane.
 
For 3d $\mathcal{N}=4$ theories, mirror symmetry involves two
or more theories with a different UV description flowing to the same
superconformal point in the IR. Mirror symmetry interchanges Coulomb and Higgs
branches of the theory, where FI  parameters are interchanged with some linear combination of mass parameters
\cite{Intriligator:1996ex}. 
The present theory is  known to be dual to a $A_{N-1}$ quiver gauge theory \cite{Intriligator:1996ex,Kapustin:2010xq}. 
Particularizing to
our model, we see that the dual theory  is a
$U(1)^{2N-1}$ quiver gauge theory with a FI parameter $2m$ and a
single mass for all hypermultiplets $-\eta/2N$. 
Our results show that
the quiver gauge theory also has a  novel type of phase
transition in the limit when the number of \textit{quiver nodes} goes to infinity.

To conclude, the partition function of supersymmetric QED$_{3}$ with the FI term
is given in terms of the conical function (\ref{dos}). It is remarkable
that this simple formula encapsulates very rich physical phenomena such as
large $N$ phase transitions, asymptotic $1/N$ expansion, the emergence of complex saddle points, nonperturbative effects and aspects of mirror symmetry.

\medskip

\begin{acknowledgments}
J.R. acknowledges financial support from Projects  FPA2013-46570 (MINECO), 2014-SGR-104 (Generalitat de Catalunya). M.T. is supported by the Fund\~{a}\c{c}ao para a Ci\^{e}ncia
e Tecnologia (program Investigador FCT IF2014), under Contract No. IF/01767/2014.

\end{acknowledgments}

\end{document}